\documentclass[conference]{IEEEtran}

\IEEEoverridecommandlockouts
\usepackage{algorithm}
\usepackage[noend]{algpseudocode}
\usepackage{listings}
\usepackage[flushleft]{threeparttable}
\usepackage{float}
\usepackage[dvipsnames]{xcolor}
\usepackage{multirow}
\usepackage{booktabs, tabularx} 
\usepackage{stfloats}
\usepackage{booktabs}
\usepackage{multicol}
\usepackage{graphicx}
\usepackage[utf8x]{inputenc}
\usepackage{multirow}
\usepackage{pifont}
\usepackage{xcolor,colortbl}
\usepackage{xspace}
\usepackage{longtable}
\usepackage{enumitem}
\usepackage{lscape}
\usepackage{url}
\usepackage[pagebackref]{hyperref}
\usepackage{amssymb}
\usepackage{pifont}

\definecolor{LightGray}{gray}{0.90}
    
\newif\ifEditMode

\EditModetrue
    
\hypersetup{pdfstartview=FitH,
  pdfpagelayout=SinglePage,
  bookmarksnumbered=true,
  bookmarksopen=true,
  colorlinks=true,
  anchorcolor=blue,
  citecolor=purple,
  linkcolor=red,
  urlcolor=blue}

\algnewcommand\algorithmicforeach{\textbf{for each}}
\algdef{S}[FOR]{ForEach}[1]{\algorithmicforeach\ #1\ \algorithmicdo}

\newcommand{\thead}[1]{\textbf{\textit{#1}}}
\newcommand{\ea}{{et al.}\xspace}

\newcommand{\mysubsection}[1]{\noindent\textbf{#1. }\ignorespaces}

\graphicspath{{./figures/}}

\def\pending#1{}

\definecolor{codegreen}{rgb}{0,0.6,0}
\definecolor{backcolour}{rgb}{0.95,0.95,0.92}
\definecolor{LightGray}{gray}{0.9}
\definecolor{Amber}{rgb}{1.0, 0.75, 0.0}

\usepackage{xcolor}
\definecolor{codegreen}{rgb}{0,0.6,0}
\definecolor{backcolour}{rgb}{0.95,0.95,0.92}
\definecolor{LightGray}{gray}{0.9}
\definecolor{Amber}{rgb}{1.0, 0.75, 0.0}

\usepackage{xcolor}
\usepackage{listings}

\begin{document}

\title{Towards a Security Stress-Test for Cloud Configurations
\thanks{This work has received funding from the European Union under the H2020 grant 952647 (AssureMOSS).
}}

\author{\IEEEauthorblockN{1\textsuperscript{st} Francesco Minna}
\IEEEauthorblockA{\textit{Vrije Universiteit Amsterdam (NL)} \\
f.minna@vu.nl}
\and
\IEEEauthorblockN{2\textsuperscript{nd} Fabio Massacci}
\IEEEauthorblockA{\textit{University of Trento (IT)}\\
\textit{Vrije Universiteit Amsterdam (NL)} \\
fabio.massacci@ieee.org}
\and
\IEEEauthorblockN{3\textsuperscript{rd} Katja Tuma}
\IEEEauthorblockA{\textit{Vrije Universiteit Amsterdam (NL)} \\
k.tuma@vu.nl}
}

\maketitle

\begin{abstract}
Securing cloud configurations is an elusive task, which is left up to system administrators who have to base their decisions on ``trial and error'' experimentations or by observing good practices (e.g., CIS Benchmarks). 
We propose a knowledge, AND/OR, graphs approach to model cloud deployment security objects and vulnerabilities. In this way, we can capture relationships between configurations, permissions (e.g., CAP\_SYS\_ADMIN), and security profiles (e.g., AppArmor and SecComp), as first-class citizens. 
Such an approach allows us to suggest alternative and safer configurations, support administrators in the study of what-if scenarios, and scale the analysis to large scale 
deployments.
We present an initial validation and illustrate the approach with three real vulnerabilities from known sources.
\end{abstract}

\begin{IEEEkeywords}
knowledge graph, AND/OR graphs, containers, security, microservices, cloud
\end{IEEEkeywords}


\section{Introduction}\label{s:introduction}

Container engines and orchestration tools, such as Docker and Kubernetes, abstract the complexity of the underlying technologies (e.g. Linux namespaces and Control Groups).
This power of abstraction reduces companies' time to market, by making it easier to deploy and share applications. Yet, it also rises new security challenges.
Indeed, it is hard to understand such hidden complexity 
in a highly dynamic environment (44\% of containers are executed for less than five minutes~\cite{sysdig22}). Misconfiguration is the most prevalent vulnerability in the cloud~\cite{nsa-cloud}, 76\% containers are running as root, and 75\% of containers are running with critical vulnerabilities~\cite{sysdig22} (\S\ref{s:running-example}).
\looseness=-1

Several tools exist for analyzing container deployments and for 
vulnerability scanning. Yet, as we illustrate in the related
section (\S\ref{s:related-work}), most `vulnerability scanners' are actually simply
checkers of bills of (software materials). When actual configurations
are checked (e.g. the Checkov tool) they are checked against best practices (i.e. CIS Benchmarks).
Such analysis reports
rules that are not met but cannot 
do a more sophisticated what-if analysis of what can be changed 
while maintaining a satisfying secure configuration.

\begin{figure*}[tb]
    \centering
    \includegraphics[width=18cm]{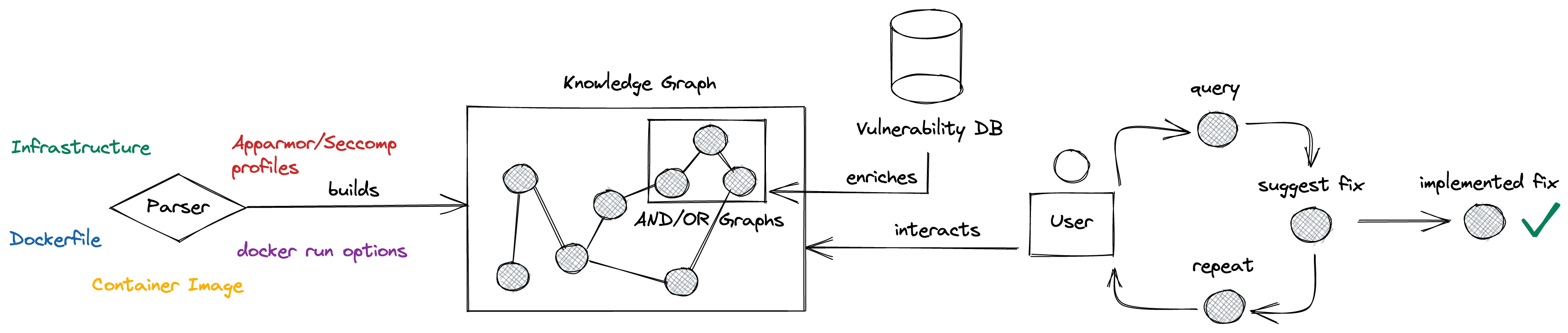}

    \caption{The workflow of the stress-test methodology on Docker containers.}
    \label{fig:sol-design}
\end{figure*}

To fill this gap, we propose to \emph{use a knowledge graphs where objects, permissions, capabilities and
other security features (e.g. AppArmor for Docker) are first class citizens} of the model.
To this end, we build such knowledge graphs and contribute with search algorithms implemented in Python and Neo4J (\S\ref{s:design}) to identify vulnerabilities and their fixes
(\S\ref{s:fix}).
Fig.~\ref{fig:sol-design} shows the workflow steps of the 
stress-test on Docker containers, namely, (i) building the 
knowledge graph from configuration files, (ii) 
enriching the graph with known vulnerabilities, and (iii), 
testing and fixing the system 
by interacting with the user.

The idea is to construct with this methodology the equivalent of a stress-test\footnote{``the stress test for banks is used to assess their resilience to adverse economic and market developments and contributes to the assessment of systemic risk''~\cite{stress-test}.} for cloud configuration.
The methodology will allow administrators to explore what-if scenarios whenever a new vulnerability is discovered and better understand the trade-offs made between relaxing security requirements and functionality loss.

We present an initial validation by illustrating the proposed approach in \S\ref{s:result} and discuss the results and future work in \S\ref{s:discussion}.
We present the concluding remarks in\S\ref{s:conclusion}.
%

\section{Simple Attack Scenarios}
\label{s:running-example}

To illustrate the problem faced by administrators in charge of cloud 
deployment, we describe two recent attack scenarios and their implications for 
containers.
In 2019, a researcher from Google posted on Twitter a proof of concept (PoC) 
to escape from privileged Docker containers, by exploiting the release\_agent 
of control groups (cgroups)~\cite{docker-escape}. 
Listing~\ref{listing:config} shows 
a possible vulnerable configuration. This configuration adds
system administrator capabilities to the default
root user of Docker and tells the system to avoid
a specific AppArmor Linux profile, a solution
typically used to make sure applications can
more easily communicate across network sockets. 

\begin{lstlisting}[language=bash,caption={Vulnerable execution of the container taken from~\cite{docker-escape} \label{listing:config}},basicstyle=\footnotesize,breaklines=true,numbers=left,linewidth=\columnwidth]
 # On the host
 docker run --rm -it --cap-add=SYS_ADMIN --security-opt apparmor=unconfined ubuntu bash
 \end{lstlisting}


By creating a new cgroup, enabling cgroups notifications on release, and 
specifying the release\_agent script to be executed (once all processes within 
the same cgroup are killed) an attacker can 
execute code on 
the container's host.

There is no syntactic connection between the attack
commands 
and the Docker configuration in Listing~\ref{listing:config} so that a 
\emph{semantic signature} of the Docker configuration is needed. 

To exploit this technique, a container must either run as privileged or with a root user (by default on Docker), with CAP\_SYS\_ADMIN capability granted and mount syscall allowed. 
By foiling any (or all) preconditions the attack will fail.
For example, one could just drop all capabilities (i.e. \texttt{--cap-drop ALL}).
Yet this is hardly a side-effect free solution.
By dropping the unconfined AppArmor profile 
users will find themselves trapped with a lengthy debugging of possible calls to 
network sockets of applications that don't work and
each of which must be individually configured, debugged, and 
profiled\footnote{\url{https://ubuntu.com/server/docs/security-apparmor}}. 
%
Similarly, in March 2022, a vulnerability affecting the Linux kernel leading to privilege escalation was discovered, namely, CVE-2022-0492~\cite{cve-2022-0492}.
Again, this vulnerability exploits a vulnerable implementation of the release\_agent binary executed once all processes belonging to a cgroup are killed.
%
For this vulnerability to be exploitable, one must run a container as a root user (or with the CAP\_DAC\_OVERRIDE capability), with mount and unshare syscalls allowed; obviously, containers with more permissions (e.g. privileged or with the CAP\_SYS\_ADMIN capability) are also affected. 
The key observation is that if the configuration is changed
the vulnerable binary can stay as it is because the attack is no longer possible.

\section{Related Work}
\label{s:related-work}

\textbf{Access control and network security policies} for cloud deployment 
are well studied by providing either  access control solutions~\cite{Li2019b,Preuveneers2017,Ranjbar2017,Sun2016}
or by supporting the
automated policy generation from given specifications \cite{Fadhel18, Li2019b,zhu2021apparmor}.
Recent works also provide mechanisms for formal verification of microservice deployments \cite{Gerking2019}, or
continuous assessment methodologies
\cite{Torkura2017}.
Yet, the automatic or even computer-aided exploration and validation of configuration policies \emph{as available in 
practical deployments} (e.g.\ combinations of Docker commands) and their subsequent testing, still remain largely unexplored in the academic literature. 

\textbf{Microservice Attack Generation} Ibrahim~\ea~\cite{Ibrahim2019} investigated (theoretical) attacks on
microservices by using vulnerabilities identified in a deployment to build an attack graph of all possible attacks. Whether these attacks are actually possible has not been tested. Also this analysis cannot be readily translated to changes
into a practical configuration (e.g. drop the CAP\_SYS\_ADMIN option).

\textbf{Security industrial tools} offering different kinds of 
`vulnerability scans' 
and `static analysis' for cloud environments are also 
available on the market. Yet, while they all `formally' meet 
NIST definition
of vulnerability scanning (`a technique used to identify 
hosts attributes and associated vulnerabilities'~\cite{vuln-scanner}) their actual operation 
is not what
a security expert would intuitively expect
even from the classical \texttt{nmap} network 
scanner~\cite{lyon2008nmap} and even
less of what one would expect from a static analyzer for 
software security \cite{chess2004static}. 

Configuration analyzers (e.g. snyk~\footnote{\url{https://snyk.io/product/container-vulnerability-management/}} and Docker Bench~\footnote{\url{https://github.com/docker/docker-bench-security}}) analyze configuration files and settings and check them against common or custom security best practices and policies (e.g. compliance with the CIS Benchmarks), some of these tools only provide compliance tests.\footnote{E.g. terrascan by Accurics, built on top of OPA, and Bridgecrew, which are open source, along with CIS-CAT Pro, and Lacework which are commercial.}

Vulnerability Scanners (e.g. Clair~\footnote{\url{https://github.com/quay/clair}} and Trivy~\footnote{\url{https://github.com/aquasecurity/trivy}}) typically retrieves a 
software bill of material (SBOM) gathering all software 
packages used in an image or in configuration files and 
compares them against  vulnerabilities repositories (e.g. 
NVD\footnote{\url{https://nvd.nist.gov/}} or Debian 
Security 
Bug Tracker\footnote{\url{https://security-tracker.debian.org/tracker/}}) to check if the 
current version of a package in use is \emph{reported to 
be} vulnerable.

Such tools often generate false alerts as they are based on 
the same overcautious vulnerability reporting that is 
well known in secure software engineering for the analysis 
of vulnerabilities in  open source 
libraries~\cite{pashchenko2020vuln4real,dashevskyi2018screening}.

\textbf{Automatically suggesting a ``fix''} for a particular 
(cloud) misconfiguration is, to this day, an open problem. 
All previous tools do not suggest or provide any fix for 
detected vulnerabilities, leaving up to the end user the 
mitigation process to eliminate the alerts.
To this extent, we will use the Analytic Hierarchy 
Process (AHP)~\cite{vaidya2006analytic}, which has been 
used for decades by nearly all decision making disciplines, 
including graph theory and requirements engineering.

\textbf{Knowledge graphs for cybersecurity.} 
The use of graphs to investigate the security properties of 
cloud environments have already been investigated in the 
past. Yet, most existing approaches use knowledge graphs for 
security awareness, and intelligence sharing 
\cite{noel2016cygraph,jia2018practical}, rather then 
security analysis of configurations.
A comprehensive review can be found 
in~\cite{zhang2020review}.

The closest work is by Banse et al.~\cite{banse2021cloud} 
who build a Cloud Property Graph (CloudPG) using both 
static code analysis and an ontology; similarly to our idea, 
they propose an analysis framework to query the graph for 
security issues but focuses on bridging 
the gap between static code analysis of the application and 
its runtime environment. While the outcome is a 
sophisticated analysis of data flow leaks, this approach 
requires full access to the application code. Our approach 
complement this work by going in the opposite direction 
towards the analysis of deployment information to suggest 
alternative, safer, configurations.

\section{Building the Graphs}\label{s:design}

The methodology for capturing and analyzing cloud 
containers configuration by a knowledge graph where
undirected edges and nodes are labeled to
represent the container system objects and their relationships. 

\textbf{Build the knowledge graph.} We parse the configuration files to 
represent the infrastructure on which the container engine is running (e.g. Ubuntu 
VM), the container engine being used (e.g. Docker), the container image (e.g. 
Dockerfile), and finally, the docker run options that will be used to run the container 
(e.g. privileged, volume, and user) along with the security profiles (e.g. AppArmor and 
Seccomp files).
We then map these objects to node and their logical relations to edges in the graph 
(e.g., $Container$ (node) has capability (edge) $SYS\_ADMIN$ (node) or 
$Docker\_Engine$ (node) is using (edge) $Kernel$ (node) of version (edge) $4.9$ 
(node)).

In addition, we do not only model the current configuration (as parsed) but also 
possible configuration \textit{alternatives}. For instance, we add nodes representing 
alternative versions of the Linux kernel or Docker engine.

We choose the Neo4J graph database to represent the knowledge graph and 
vulnerabilities dataset (AND/OR graphs) and the Python language to parse 
configuration files and automatically update the graph using Neo4J APIs.

\begin{figure*}[tb]
    \centering
    \includegraphics[width=18cm]{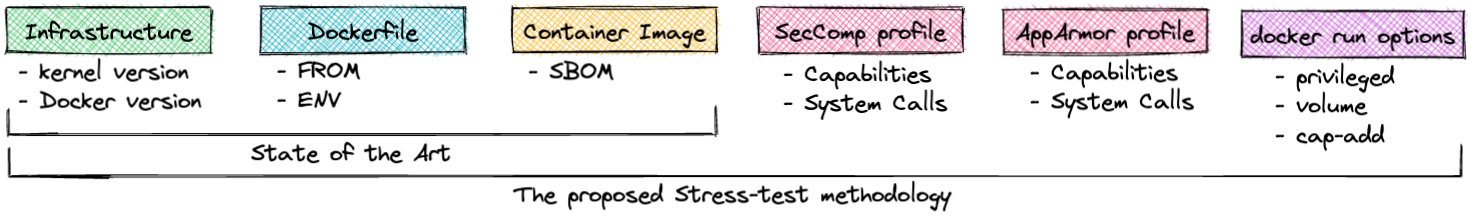}
    
    \caption{The parsed configuration files to build the knowledge graph.}
    \label{fig:config-parse}
\end{figure*}

 In order to build the knowledge graph, we 
first parse the container configurations files. 
In particular, Fig.~\ref{fig:config-parse} shows the different files that we parse (e.g., 
we also consider security profiles and docker run options), and the information 
extracted from each, to build the knowledge graph.

State of the art tools do not parse the given permissions (e.g. Linux capabilities and system calls) and the docker run options for possible misconfigurations.
For example, Dockerfile instructions can be overwritten by docker run options, thus it is important to parse all configurations available.

\textbf{Build vulnerabilities graphs} We enrich the graph with a 
dataset of known vulnerabilities and misconfigurations~\cite{containers-matrix}.

We use a two-dimensional taxonomy to classify container vulnerabilities and 
misconfigurations, slightly based on the attack taxonomy proposed in~\cite{3274720}.
%
Within the first dimension we classify the vulnerabilities in three categories 
based on the affected component, namely, \textit{Container}, \textit{Kernel}, and \textit{Engine}. 
Within the second dimension, we classify the vulnerabilities based on the 
MITRE ATT\&CK tactics and techniques (e.g. privilege escalation and network service 
scanning)~\cite{containers-matrix}.

We represent each vulnerability in the graph as an AND/OR graph, where the root is 
the final goal of the attack, the rest of the nodes represent the attack assumptions that 
must hold for an exploit.
For example, we use an OR node to represent a set of vulnerable versions (e.g. 
Linux kernel versions between $5.1$ and $5.9$) of a component that can be exploited.
Based on the running configuration (infrastructure, container engine, etc.), represented 
in the knowledge graph, we then link the latter to AND/OR graphs assumptions being 
currently used (e.g., $Container$ (node) has capability (edge) $SYS\_ADMIN$ (node) 
assumption of $CVE-xxx-xxx$ (node)).
\looseness=-1

%
The vulnerabilities dataset is represented using a JSON file; in particular, Listing~\ref{listing:vuln} shows the JSON representation of the first container escape described in \S\ref{s:running-example}.

\begin{lstlisting}[language=xml,caption={An example of a Linux kernel vulnerability representation.\label{listing:vuln}}, basicstyle=\footnotesize,breaklines=true,numbers=left,linewidth=\columnwidth] 
"CVE-2022-0492": [{
    "CVSS_v3": "7.8",
    "mitre_tactic": "priviledge_escalation",
    "mitre_technique": "escape_to_host",
    "pre_conditions": [
        {
            "max_kernel_version": "5.17 rc3",
            "capability": "CAP_DAC_OVERRIDE",
            "syscall": "mount, unshare"
        }
    ],
    "post_conditions": [
        {
            "impact": "priviledge_escalation"
        }
    ]
}]       
\end{lstlisting}

\section{Searching for a Fix} \label{s:fix}

\textbf{Analytic Hierarchy Process (AHP)}. 
The AHP is a multiple criteria decision-making methodology. In AHP, a hierarchy of solution alternatives is created, then the user is asked to provide weights to each solution alternative, and finally, pair-wise alternatives are compared until the optimal solution is obtained~\cite{VAIDYA20061}.
Before evaluating the presence of vulnerabilities within the graph, we interact with the 
end-user to rank the list of possible fixes that can mitigate the attacks.
For instance, we may ask a user to assign a preference  
(with $1$ least favorite and $9$ most favorite) to run a container as not root 
or assign a preference to remove specific capabilities of the container.
In practice this means ordering the following list of possible fixes: 
(version\_upgrade, not\_privileged, not\_root, not\_capability, 
not\_syscall, read\_only\_fs, no\_new\_priv).
Similar dimensions can then be used on different types of fixes. 
This ranking will help us in suggesting the most effective fix that, at the same time, the 
user is also most willing to accept and implement.
\looseness=-1

\textbf{Test \& Fix for vulnerabilities.} Finally, we test (by querying the graph) and fix (by suggesting and eventually implementing security fixes) the system. To this aim, we rely on the user to accept the best alternative configurations (based on the provided weights). An attack is mitigated by implementing a  fix (invalidating one attacks' assumption) which is reflected in the graph by removing or adding an edge.

Algorithm~\ref{alg:query} shows the procedure to stress-test a cloud system by unveiling the presence of one vulnerability.
The steps of the algorithm are (1) parsing the configuration files and building the knowledge graph, (2) adding and link the vulnerability AND/OR graph, (3) fetching user preferences, and finally (4) suggesting the best fix.
To this aim, the algorithm iteratively suggests security fixes to mitigate the vulnerability assumptions, following the user preferences. 
Eventually, the algorithm can automatically accept the most preferred fix.

\begin{algorithm}

    \caption{Stress-test a cloud system for one vulnerability.}
    \label{alg:query}
    
\begin{algorithmic}[1]
\Require Configuration files
\Require Vulnerability
\Ensure Safe configuration
\State Build Knowledge graph;
\State Enrich the graph with the vulnerability;
\State Get list of user preferred fixes;
\State \Call{get\_preferred\_fix}{$vuln\_query$}

\Function{get\_preferred\_fix}{$vul\_query$}
  \While{$vuln\_query == true$}
    \State Collect vulnerability assumptions;
    \State Order assumptions by AHP;
    \Repeat
    \State Suggest assumption to remove;
    \If{$user\_accepts\_fix$}
        \State remove edge;
     \EndIf
    \Until{$user\_not\_happy$}
    \EndWhile
\EndFunction
\end{algorithmic}
\end{algorithm}

The configuration of more realistic cloud deployments may contain several 
vulnerabilities. In this case, the problem requires finding a globally optimal solution.
Therefore, local fixes must be additionally weighted in combination with other 
misconfiguration fixes. Currently, our algorithm relies on user input to determine such weights based on AHP. Our implementation could be extended in the future to find a set of globally optimal fixes.

In some cases, the user can also decide to intentionally keep a few low-risk 
vulnerabilities to avoid trading off the performance of the system.
In such cases, although the vulnerability query will still be satisfied, this approach may 
help move towards a "safer" configuration state (e.g. by running a container with all 
and only needed capabilities and system calls, instead of privileged).

\looseness=-1

We provide Neo4J queries examples in the \textit{Preliminary Validation} paragraph of \S\ref{s:result}.

\section{Preliminary Evaluation}\label{s:result}

We investigated the cost of storing the representation of container configurations (incl. 
configuration alternatives) and the memory requirements for storing vulnerabilities in the knowledge graph.
In addition, we illustrate the feasibility of the proposed approach on three vulnerabilities.

\mysubsection{The cost of storing containers} Tab.~\ref{tab:cont-cost} shows the cost, in terms of number of nodes and edges, of storing a container image and several containers (using the Docker default security profile, with a subset of capabilities and system calls allowed).
Specifically, we use one node to represent the host virtual machine, $50$ nodes to represent the versions of Linux kernel (from version $3.9$, released after March 20 2013, when Docker was released), one node to represent the Docker engine, $132$ nodes to represent the versions of \texttt{Docker}, $83$ nodes to represent the versions of \texttt{containerd}, and $18$ nodes to represent the versions of \texttt{runc}.
In addition, we use $41$ nodes to represent Linux capabilities and $364$ nodes to represent Linux system calls.
\looseness=-1

This amounts to $691$ nodes and $6$ relationships (first row in Tab.~\ref{tab:cont-cost}).
Additional nodes are used to represent container permissions, security profiles, and 
Docker run options; for example, custom security profiles (e.g., user-defined AppArmor profiles), may require additional nodes and relationships corresponding to the granted capabilities and system calls, as well as relevant Docker run options (e.g., \textit{-{}-user} and \textit{-{}-volume}).

{\renewcommand{\arraystretch}{1} 
\begin{table}[ht]
  \caption{The cost of storing container images and (default) configurations into the knowledge graph.}
  \label{tab:cont-cost}

  \begin{center}
    \begin{tabular}{l r r}
    \toprule
    \centering \thead{\texttt{Object/s}}
    & \thead{\texttt{\#\_Nodes}}
    & \thead{\texttt{\#\_Edges}} \tabularnewline
    
    \midrule
    \textit{Initialization} & $691$ & $6$ \\
    
    \midrule
    $1$ Image & $693$ & $8$ \\
    
    \midrule
    $1$ Container (default) & $702$ & $349$ \\
    
    \midrule
    $100$ Container (default) & $999$ & $1.339$ \\
    
    \midrule
    $1.000$ Container (default) & $3.699$ & $10.339$ \\
    
    \bottomrule

    \end{tabular}
  \end{center}
  
\end{table}
}

In the previous table, we considered all containers to be instantiated from different images, thus with two additional nodes and relationships.
Overall, the number of nodes and edges in the knowledge graph does not grow exponentially.
This is promising as it would allow us defining efficient algorithms to suggest the best security fixes.

Tab.~\ref{tab:cont-config} shows different Docker run options with the corresponding number of edges; the number of nodes in the graph remains the same.

{\renewcommand{\arraystretch}{1} 
\begin{table}[ht]
  \caption{List of possible Docker run options with the corresponding number of edges.}
  \label{tab:cont-config}

  \begin{center}
    \begin{tabular}{p{6.3cm} c}
    \toprule
    \centering \thead{\texttt{Docker run options}}
    & \thead{\texttt{\#\_Edges}} \tabularnewline
    
    \midrule
    \texttt{--cap-drop ALL --cap-add NET\_BIND\_SERVICE} & $335$ \\
    
    \midrule
    \texttt{--cap-add NET\_ADMIN --security-opt apparmor=unconfined} & $348$ \\
    
    \midrule
    \texttt{--privileged} & $420$ \\

    \bottomrule

    \end{tabular}
  \end{center}
  
\end{table}
}

Additional configurations exist, as a combination of adding and removing capabilities, granting or denying system calls (within \textit{AppArmor} or \textit{SecComp}), not using the root user or a read-only filesystem.\\
\mysubsection{The cost storing vulnerabilities} Tab.~\ref{tab:vulnerabilities} presents the dataset of vulnerabilities we collected from different sources, along with the number of nodes and edges needed to represent them.
The number of nodes and edges in the table corresponds to the AND/OR graph representing each vulnerability; for example, in the graph there is a node presenting each Linux kernel version, and for a kernel vulnerability, there will be an edge between the vulnerability and each vulnerable version.
We classify vulnerabilities in three categories, namely, \textit{container profile misconfigurations}, \textit{Linux kernel bugs}, and \textit{container engine vulnerabilities}, as suggested in~\cite{9652631}.
Within each category, we have a subcategory representing the impact of each vulnerability or attack based on the MITRE ATT\&CK tactic and techniques for containers~\cite{containers-matrix}.
For the validation of our stress-test approach, we retrieved a set of vulnerabilities belonging to each category.
In particular, for \textit{container profile misconfigurations} we retrieved $6$ escaping attacks, for \textit{Linux kernel bugs} we retrieved $24$ CVEs (only 10 shown in Tab.~\ref{tab:vulnerabilities}), mostly from~\cite{3274720} and, finally, for \textit{container engine vulnerabilities}, considering only Docker (or Docker subcomponents, such as \textit{runc} and \textit{containerd}), we retrieved $6$ CVEs from~\cite{9652631}.
In total, our dataset contains $36$ vulnerabilities.

{ 
\aboverulesep = 0mm
\belowrulesep = 0mm
\renewcommand{\arraystretch}{1.1} 

\begin{table}[ht]
  \caption{Information about the vulnerabilities dataset, divided by categories, together with the number of nodes and edges.}
  \begin{center}
    \begin{tabular}{c c c c c}
    \toprule
    \centering \thead{\texttt{Categ.}}
    & \thead{\texttt{Sub-cat.}}
    & \thead{\texttt{CVE}} 
    & \thead{\texttt{\#\_Nodes}}
    & \thead{\texttt{\#\_Edges}} \tabularnewline
    
    \midrule
    \rowcolor{YellowGreen}
    \multirow{6}{1cm}{Container misc.} & \multirow{6}{1cm}{Escape to host} & Cgroup escape & $8$ & $12$ \\ 
    & & Cap. SYS\_MODULE & $6$ & $7$ \\
    & & Kernel fs /sys & $2$ & $2$ \\
    & & Host devices & $7$ & $11$ \\
    & & Host ~/root & $4$ & $6$ \\
    & & Docker socket & $4$ & $6$ \\
    
    \midrule
    \multirow{10}{1cm}{Kernel bugs} & & CVE-2022-0847 & $11$ & $19$ \\
    \rowcolor{YellowGreen}
    & & CVE-2022-0492 & $58$ & $62$ \\
    & & CVE-2022-0185 & $22$ & $25$ \\
    & & CVE-2020-14386 & $44$ & $44$ \\
    & & CVE-2017-7308 & $47$ & $47$ \\
    & Exploitation & CVE-2017-5123 & $10$ & $10$ \\
    & & CVE-2016-8655 & $37$ & $40$ \\
    & & CVE-2016-4997 & $20$ & $20$ \\
    & & CVE-2017-6074 & $23$ & $23$ \\
    & & CVE-2017-1000112 & $27$ & $27$ \\
    
    \midrule
    \multirow{6}{1cm}{Docker vuln.} & Code injection & CVE-2019-14271 & $9$ & $14$ \\
    & Exploitation & CVE-2020-15257 & $106$ & $109$ \\
    & \multirow{3}{1cm}{Escape to Host} & CVE-2016-9962 & $27$ & $29$ \\
    & & CVE-2018-15664 & $66$ & $66$ \\
    & & CVE-2019-5736 & $299$ & $300$ \\
    \rowcolor{YellowGreen}
    & DoS & CVE-2020-13401 & $221$ & $224$ \\
    
    \bottomrule
    \end{tabular}
  \end{center}
\end{table}
\label{tab:vulnerabilities}
}

\mysubsection{Vulnerabilities Queries} To illustrate the feasibility of our approach, we performed an initial evaluation on three vulnerabilities, namely, \textit{Cgroup escape}, \textit{CVE-2022-0492}, and \textit{CVE-2020-13401} (green rows in Tab.~\ref{tab:vulnerabilities}).
To the best of our knowledge, an existing algorithm to traverse AND/OR graph in Neo4j does not yet exist.
Therefore, for our initial validation, we manually check whether a vulnerability is exploitable or not.
In future work, we plan to implement such an algorithm to automatically check the presence of vulnerabilities.
Listing~\ref{listing:vuln-query1} shows a Neo4J query to check whether a certain 
property (e.g., permission, capability, or system call) belongs to a container 
or not.
\looseness=-1

\begin{lstlisting}[caption={Neo4J query to check a
container relationship\label{listing:vuln-query1}}, 
basicstyle=\footnotesize,breaklines=true,numbers=left,linewidth=\columnwidth] 
MATCH (c:Container {name: "Nginx"}) 
MATCH (p:Permissions {name: "Privileged"}) 
RETURN EXISTS( (c)-[:HAS]-(p)    
\end{lstlisting}

Based on the current implementation, as future work, we plan to test this approach on all $36$ vulnerabilities with different configurations (different security profiles, docker run options, etc.) and possibly, on different container engines.
By doing so, we can evaluate more extensively the results of stress-tests and explore the support of what-if scenarios; for example, we can claim the most effective mitigation against most attacks but also evaluate the most dangerous configurations.
%
Finally, at the end of each run, the number of removed edges (attacks assumptions) could be used as an indicator of the resilience to the stress test and can be used as a measure of quality in the context of a more comprehensive validation.
\looseness=-1

\section{Discussion}\label{s:discussion}

This approach might also be suitable to build attack 
paths in more complex environments (e.g. Kubernetes 
clusters).

Efficient graph algorithms could be used to cast and then 
address a number of security issues: 
reachability (e.g. given 
the current configuration, whether the attacker can reach a 
new state), minimum weight traversal (e.g. the 
easiest exploitation path), and vertex cuts (e.g. suggesting a 
policy or configuration change that invalidates an attack).

To have a dynamic picture, one could run the stress-test at 
a given time interval and evaluate the output at each time; 
for example, we could evaluate the number of changes in 
the graph (new nodes or edges) as a risk analysis of the 
the current configuration, or the number of changes needed to 
get to a "safe" state.
\looseness=-1

We plan to extend our implementation and support more 
container engines (e.g. Podman and CRI-O) as well as 
infrastructure-as-code tools (e.g. Vagrant and Terraform) 
integrate the implementation into a container orchestration 
tool (e.g. in Kubernetes as an admission controller).

\section{Conclusion}\label{s:conclusion}

In this paper, we proposed a stress-test approach for cloud configurations, by using a knowledge graph to represent deployments and permissions, and AND/OR graphs to represent vulnerabilities.
We described our solution along with algorithms to unveil vulnerabilities and semi-automatically suggest security fixes to mitigate them; we also described a first implementation and evaluation on Docker containers.
We believe this is the first step forward towards a dynamic and continuous evaluation mechanism for risk analysis in cloud environments while suggesting security fixes.
In the future, more validation on scalability and query performance is needed, especially with hundreds or thousands of containers.


\section*{Acknowledgments}

This work has been partially supported by the European Union under the H2020 grant 952647 (AssureMOSS).

\bibliographystyle{IEEEtran}

\end{document}